\documentclass{article}
\usepackage{amsmath,amssymb,amsfonts}
\usepackage{geometry}
\usepackage{multirow}

\usepackage[backref=page]{hyperref}
\hypersetup{
    unicode=false,          
    pdftoolbar=true,        
    pdfmenubar=true,        
    pdffitwindow=true,      
    pdftitle={A Note on Element-wise Matrix Sparsification via a Matrix-valued Bernstein Inequality},    %
    pdfauthor={Petros Drineas, Anastasios Zouzias},     
    pdfsubject={Matrix Sparsificaiton},   
    pdfcreator={Petros Drineas and Anastasios Zouzias},   
    pdfproducer={Petros Drineas}, 
    pdfkeywords={probability,Bernstein, algorithm, matrix-valued, matrix sparsification}, 
    pdfnewwindow=true,      
    colorlinks=false,       
    linkcolor=red,          
    citecolor=green,        
    filecolor=magenta,      
    urlcolor=cyan          
}

\usepackage[ruled,boxed]{algorithm}
\usepackage{algorithmic}

\renewcommand*{\backref}[1]{}
\renewcommand*{\backrefalt}[4]{%
	\ifcase #1 %
		(Not cited) %
	\or
		(Cited on page~#2)%
	\else 
		(Cited on pages~#2)%
	\fi
}

\newcommand{\abs}[1]{\left|#1\right|}
\newcommand{\R}{\mathbb{R}}
\newcommand{\Prob}{\mathbb{P}}
\newcommand{\E}{\mathbb{E}}
\newtheorem{thm}{Theorem}
\newtheorem{lem}{Lemma}
\newtheorem{cor}{Corollary}
\newenvironment{Proof}{\noindent {\em Proof:}}{\\\hspace*{\fill}\mbox{$\diamond$}}

\newcommand{\norm}[1]{\ensuremath{\left\|#1\right\|_2}}

\newcommand{\frobnorm}[1]{\ensuremath{\left\|#1\right\|_{\text{\rm F}}}}
\newcommand{\sr}[1]{\ensuremath{\mathrm{\textbf{\footnotesize sr}}\left(#1\right)}}

\newcommand{\Id}{\mathbf{I}}

\newcommand{\zeromtx}{\mathbf{0}}

\begin{document}

\title{A Note on Element-wise Matrix Sparsification via a Matrix-valued Bernstein Inequality}
\author{
Petros Drineas\thanks{Department of Computer Science, Rensselaer Polytechnic Institute, \href{mailto:drinep@cs.rpi.edu}{drinep@cs.rpi.edu}}
\and
Anastasios Zouzias\thanks{Department of Computer Science, University of Toronto, \href{mailto:zouzias@cs.toronto.edu}{zouzias@cs.toronto.edu }}}
\maketitle

\begin{abstract}
\noindent Given a matrix $A \in \R^{n \times n}$, we present a simple, element-wise sparsification algorithm that zeroes out all sufficiently small elements of $A$ and then retains some of the remaining elements with probabilities proportional to the square of their magnitudes. We analyze the approximation accuracy of the proposed algorithm using a recent, elegant non-commutative Bernstein inequality, and compare our bounds with all existing (to the best of our knowledge) element-wise matrix sparsification algorithms.
\end{abstract}
%
%
\section{Introduction}
Element-wise matrix sparsification was pioneered by Achlioptas and McSherry~\cite{AM01,AM07}, who described sampling-based algorithms to select a small number of elements from an input matrix $A \in \R^{n \times n}$ in order to construct a sparse sketch $\widetilde A \in \R^{n \times n}$, which is close to $A$ in the operator norm. Such sketches were used in approximate eigenvector computations~\cite{AM01,AHK06,AM07}, semi-definite programming solvers~\cite{AHK05,Asp09}, and matrix completion problems~\cite{CR09,CT09}. Motivated by their work, we present a simple matrix sparsification algorithm that achieves the best known upper bounds for element-wise matrix sparsification.

Our main algorithm (Algorithm~1) zeroes out ``small'' elements of $A$ and randomly samples the remaining elements of $A$ with respect to a probability distribution that favors ``larger'' entries.
\begin{algorithm}
\centerline{\caption{Matrix Sparsification Algorithm}}
\begin{algorithmic}[1]
\STATE \underline{\textbf{Input:}} $A \in \R^{n \times n}$, accuracy parameter $\epsilon >0$.
\STATE \textbf{Let} $\widehat A = A$ and \textbf{zero-out} all entries of $\widehat A$ that are smaller (in absolute value) than $\epsilon/2n$.
\STATE \textbf{Set} $s$ as in Eqn.~\eqref{eqn:sfinal}.
\STATE \textbf{For} $t = 1\ldots s$ (i.i.d. trials with replacement) \textbf{randomly sample} indices $(i_t, j_t) $ (entries of $\widehat{A}$), with
\[\Prob\left( (i_t, j_t) =  (i,j)\right)\ =\ p_{ij}, \quad \mbox{where }p_{ij}:=\widehat A_{ij}^2/\frobnorm{\widehat A}^2 \mbox{for all } (i,j) \in [n]\times[n].\]
%
\STATE \underline{\textbf{Output:}} \[\widetilde{A} = \frac{1}{s}\sum_{t=1}^s \frac{\widehat A_{i_t j_t}}{p_{i_t j_t}} e_{i_t} e_{j_t}^T \in \R^{n \times n}.\]
\end{algorithmic}
\end{algorithm}
In Algorithm~$1$, we let $e_1,e_2,\ldots,e_n \in \R^n$ denote the standard basis vectors for $\R^n$ (see Section~\ref{sxn:notation} for more notation). Our sampling procedure selects $s$ entries from $A$ (note that $\widehat A$ from the description of Algorithm~$1$ is simply $A$, but with elements less than or equal to $\epsilon/(2n)$ zeroed out) in $s$ independent, identically distributed (i.i.d.) trials with replacement. In each trial, elements of $A$ are retained with probability proportional to their squared magnitude. Note that the same element of $A$ could be selected multiple times and that $\widetilde A$ contains at most $s$ non-zero entries. Theorem~\ref{thm::main} is our main quality-of-approximation result for Algorithm $1$ and achieves sparsity bounds proportional to $\frobnorm{A}^2$.
\begin{thm} \label{thm::main}
Let $A \in \R ^{n \times n}$ be any matrix, let $\epsilon >0 $ be an accuracy parameter, and let $\widetilde{A}$ be the sparse sketch of $A$ constructed via Algorithm 1. If
\begin{equation}\label{eqn:sfinal}
s = \frac{28n \ln\left(\sqrt{2}n\right)}{\epsilon^{2}}\frobnorm{A}^2,
\end{equation}
then, with probability at least $1-n^{-1}$,
$$
\norm{A - \widetilde{A}} \leq \epsilon.
$$
$\widetilde{A}$ has at most $s$ non-zero entries and the construction of $\widetilde{A}$ can be implemented in one pass over the input matrix $A$ (see Section~\ref{sxn:onepass}).
\end{thm}
We conclude this section with Corollary~\ref{cor:relativeerror}, which is a re-statement of Theorem~\ref{thm::main} involving the \emph{stable rank} of $A$, denoted by $\sr{A}$ (recall that the stable rank of any matrix $A$ is defined as the ratio $\sr{A} := \frobnorm{A}^2/\norm{A}^2$, which is upper bounded by the rank of $A$). The corollary guarantees relative error approximations for matrices of -- say -- constant stable rank, such as the ones that arise in~\cite{Rec09,CT09}.
\begin{cor}\label{cor:relativeerror}
Let $A \in \R ^{n \times n}$ be any matrix and let $\varepsilon >0 $ be an accuracy parameter. Let $\widetilde{A}$ be the sparse sketch of $A$ constructed via Algorithm 1 (with $\epsilon = \varepsilon\norm{A}$). If
$s = 28n\sr{A} \ln\left(\sqrt{2}n\right) / \varepsilon^{2},$
then, with probability at least $1-n^{-1}$,
\[
\norm{A - \widetilde{A}} \leq \varepsilon\norm{A}.
\]
\end{cor}
It is worth noting that the sampling algorithm implied by Corollary~\ref{cor:relativeerror} can not be implemented in one pass, since we would need a priori knowledge of the spectral norm of $A$ in order to implement Step $2$ of Algorithm $1$.
%
%
\section{Related Work}\label{sxn:relatedwork}
In this section (as well as in Table~\ref{table:summary}), we present a head-to-head comparison of our result with all existing (to the best of our knowledge) bounds on matrix sparsification. In~\cite{AM01,AM07} the authors presented a sampling method that requires in \emph{expectation} $16 n \frobnorm{A}^2 / \epsilon^2  + 8^4 n\log^4 n$ non-zero entries in $\widetilde{A}$ in order to achieve an accuracy guarantee $\epsilon$ with a failure probability of at most $e^{-19\log^4 n}$. Compared with our result, their bound holds only when $\epsilon > 4\sqrt{n} \cdot \max_{i,j}|A_{ij}|$ and, in this range, our bounds are superior when $\frobnorm{A}^2 / (\max_{i,j}|A_{ij}|)^2=o(n\log^3 n)$. It is worth mentioning that the constant involved in~\cite{AM01,AM07} is two orders of magnitude larger than ours and, more importantly, that the results of~\cite{AM01,AM07} hold only when $n\geq 700 \cdot 10^6$.

In~\cite{GT09}, the authors study the $\|\cdot\|_{\infty \rightarrow 2}$ and $\|\cdot\|_{\infty \rightarrow 1}$ norms in the matrix sparsification context and they also present a sampling scheme analogous to ours. They achieve (in expectation) a sparsity bound of $R n\frobnorm{A}^2 \max_{i,j}|A_{ij}|  /\epsilon^2$  when $\epsilon \geq \sqrt{nR}\max_{i,j}|A_{ij}|$; here $R=\max_{ij} {|A_{ij}|} / \min_{A_{ij}\neq 0} |A_{ij}|$. Thus, our results are superior (in the above range of $\epsilon$) when $R\cdot \max_{i,j}|A_{ij}| =\omega( \log n)$.

It is harder to compare our method to the work of~\cite{AHK06}, which depends on the $\sum_{i,j=1}^n \abs{A_{ij}}$. The latter quantity is, in general, upper bounded only by $n \frobnorm{A}$, in which case the sampling complexity of~\cite{AHK06} is much worse, namely $O(n^{3/2}\frobnorm{A}^2/\epsilon)$.
%
%
Finally, the recent bounds on matrix sparsification via the non-commutative Khintchine's inequality in~\cite{drineas:sparsification_via_khintchine} are inferior compared to ours in terms of sparsity guarantees by at least $O(\ln^2 ( n /\ln^2 n))$. However, we should mention that the bounds of~\cite{drineas:sparsification_via_khintchine} can be extended to multi-dimensional matrices (tensors), whereas our result does not generalize to this setting; see~\cite{drineas:tensor_sparsification} for details.
\begin{table}[ht]
{\small
\centering
    \begin{tabular}{ | c | c | c | c | c |}
	\hline
    \multicolumn{5}{|c|}{} \\
	\multicolumn{5}{|c|}{\underline{\textbf{Comparison with Prior Results}}} \\
    \multicolumn{5}{|c|}{} \\
	\hline\hline
	\multirow{2}{*}{} & & & & \\
	\textbf{Sparsity of $\widetilde{A}$} &  & \textbf{Failure}  & \textbf{Citation}   & \textbf{Comments} \\
	 &  & \textbf{Probability}  & & \\
	\hline
	\hline
	\multirow{2}{*}{} & & & & \\
	$16 n \frobnorm{A}^2 / \epsilon^2  + 8^4 n\log^4 n$    & Expected & $e^{-19\log^4 n}$ &   \cite{AM07}  & $\epsilon > 4 \sqrt{n}\cdot b$ \\
	    & & & & $n\geq 700\cdot 10^6$ \\
	\hline
%
%
%
%
   \multirow{2}{*}{}  & & & & \\
	$R\cdot b \cdot n \frobnorm{A}^2 /\epsilon^2$   & Expected & $e^{-\Omega(R\cdot n ) }$ &   \cite{GT09}  & $\epsilon > c_1\sqrt{n\cdot R} \cdot b,\ n\geq 1$ \\ \hline
%
%
%
%
%
%
	\multirow{2}{*}{} & & & & \\
	$c_2 n \log^2 (\frac{n}{\log^2 n})\log n \frobnorm{A}^2/\epsilon^2$  & Expected  & $1/n$ &   \cite{drineas:sparsification_via_khintchine}  & $\epsilon>0,\ n\geq 300$, \\
	& & & & $c_2 \leq 45^2$ \\
	\hline
	\multirow{2}{*}{} & & & & \\
	$c_3 n \log^3 n \frobnorm{A}^2/\epsilon^2$   & Expected & $1/n$ &   \cite{drineas:tensor_sparsification}  & $\epsilon>0,\ n\geq 300$ \\
    & & & & \small{Extends to tensors} \\ \hline
    	\multirow{2}{*}{} & & & & \\
	$c_4 \sqrt{n}\sum_{ij}|A_{ij}| / \epsilon $   & Exact & $e^{-\Omega(n)}$ &   \cite{AHK06}  & $\epsilon >0$, $n\geq 1$\\
	\hline
	\multirow{2}{*}{} & & & & \\
	$28n \ln\left(\sqrt{2}n\right)\frobnorm{A}^2/ \epsilon^{2}$ & Exact & $1/n$ &  {\small Theorem~\ref{thm::main}}  & $\epsilon>0,\ n\geq 1$\\
	\hline
\end{tabular}}
\caption{Summary of prior work in matrix sparsification results. Given a matrix $A\in{\R^{n\times n}}$ and an accuracy parameter $\epsilon > 0$, we seek a sparse $\widetilde{A}\in{\R^{n\times n }}$ such that $\norm{A-\widetilde{A}} \leq \epsilon$. The first column indicates the number of non-zero entries in $\widetilde{A}$, whereas the second column indicates whether this number is exact or simply holds in expectation. In terms of notation, we let $b$ denote the $\max_{i,j}|A_{ij}|$ and $R$ denote the $\max_{ij} {|A_{ij}|} / \min_{A_{ij}\neq 0} |A_{ij}|$. Finally, $c_1,c_2,c_3,c_4$ denote unspecified constants.}\label{table:summary}
\end{table}
\section{Background}
\subsection{Notation}\label{sxn:notation}
We let $[n]$ denote the set $\{1,2,\ldots,n\}$. We will use the notation $\Prob\left(\cdot\right)$ to denote the probability of the event in the parentheses and $\E\left( X\right)$ to denote the expectation of a random variable $X$. When $X$ is a matrix, $\E\left( X\right)$ denotes the element-wise expectation of each entry of $X$. For a matrix $X \in \R^{n \times n}$, $X^{(j)}$ will denote the $j$-th column of $X$ as a column vector and, similarly, $X_{(i)}$ will denote the $i$-th row of $X$ as a row vector (for any $i$ or $j$ in $[n]$). The Frobenius norm $\frobnorm{X}$ of the matrix $X$ is defined as
$\frobnorm{X}^2 = \sum_{i,j=1}^n X_{ij}^2,$
and the spectral norm $\norm{X}$ of the matrix $X$ is defined as
$\norm{X} = \max_{\norm{y}=1} \norm{X y}.$
For two symmetric matrices $X,Y$ we say that $Y \succeq X$ if and only if $Y-X$ is a positive semi-definite matrix. Finally, $\Id_n$ denotes the identity matrix of size $n$ and $\ln x$ denotes the natural logarithm of $x$.
\subsection{Implementing the Sampling in one Pass over the Input Matrix}\label{sxn:onepass}
We now discuss the implementation of Algorithm~1 in one pass over the input matrix $A$. Towards that end, we will leverage (a slightly modified version of) Algorithm \textsc{Select} (p. 137 of~\cite{DKM06a}).
\begin{algorithm}
\centerline{\caption{One-pass \textsc{Select} algorithm}}
\begin{algorithmic}[1]
\STATE \underline{\textbf{Input:}} $A_{ij}$ for all $(i,j)\in [n]\times [n]$, arbitrarily ordered and $\epsilon >0$.
\STATE $N=0$.
\STATE \textbf{For all} $(i,j)\in [n]\times [n]$ \textbf{such that} $A_{ij}^2 > \frac{\epsilon^2}{4n^2}$
\begin{itemize}
\item $N = N + A_{ij}^2$.
\item \textbf{Set} $(I,J)=(i,j)$ and $S = A_{ij}$ \textbf{with probability} $\frac{A_{ij}^2}{N}$.
\end{itemize}
\STATE \underline{\textbf{Output:}} Return $(I,J)$, $S$ and $N$.
\end{algorithmic}
\end{algorithm}
\noindent We note that Step $3$ essentially operates on $\widehat A$. Clearly, in a single pass over the data we can run in parallel $s$ copies of the \textsc{Select} Algorithm (using a total of $O(s)$ memory) to effectively return $s$ independent samples from $\widehat A$. Lemma~$1$ (page $136$ of~\cite{DKM06a}, note that the sequence of the $A_{ij}^2$'s is all-positive) guarantees that each of the $s$ copies of \textsc{Select} returns a sample satisfying:
\[\Prob \left( (i_t,j_t) = (i, j)  \right)\ =\ \frac{\widehat A_{ij}^2}{\sum_{i,j=1}^n \widehat A_{ij}^2} = \frac{\widehat A_{ij}^2}{\frobnorm{\widehat A}^2},\quad \mbox{for all }t=1,\dots ,s.\]
Finally, in the parlance of Step $5$ of Algorithm~$1$, $(i_t, j_t)$ is set to $(I,J)$ and $p_{i_tj_t}$ is set to $S^2/N$ for all $t \in [s]$.
\section{Proof of Theorem~\ref{thm::main}}
The proof of Theorem~\ref{thm::main} will combine Lemmas~\ref{lem:lem1} and~\ref{lem:lem4} in order to bound $\norm{A - \widetilde A}$ as follows:
\begin{eqnarray*}
	\norm{A-\widetilde A} = \norm{A-\widehat A + \widehat A - \widetilde A} \leq \norm{A-\widehat A} + \norm{\widehat A - \widetilde A} \leq \epsilon/2+\epsilon/2 = \epsilon.
\end{eqnarray*}
The failure probability of Theorem~\ref{thm::main} emerges from Lemma~\ref{lem:lem4}, which fails with probability at most $n^{-1}$ for the choice of $s$ in Eqn.~(\ref{eqn:sfinal}). The proof of Lemma~\ref{lem:lem4} will involve an elegant matrix-valued Bernstein bound proven in~\cite{Rec09}. See also~\cite{chernoff:matrix_valued:Bernstein:Gross} or~\cite[Theorem~2.10]{chernoff:matrix_valued:Tropp} for similar bounds.
\subsection{Bounding $\norm{A - \widehat A}$}
\begin{lem}\label{lem:lem1}
Using the notation of Algorithm~1, $\norm{A - \widehat A} \leq \epsilon/2$.
\end{lem}
\begin{Proof}
Recall that the entries of $\widehat A$ are either equal to the corresponding entries of $A$ or they are set to zero if the corresponding entry of $A$ is (in absolute value) smaller than $\epsilon/(2n)$. Thus,
$$\norm{A-\widehat A}^2 \leq \frobnorm{A-\widehat A}^2 = \sum_{i,j=1}^n \left(A - \widehat{A}\right)_{ij}^2 \leq \sum_{i,j=1}^n \frac{\epsilon^2}{4n^2}\leq \frac{\epsilon^2}{4}.$$
\end{Proof}
\subsection{Bounding $\norm{\widehat A - \widetilde A}$}
In order to prove our main result in this section (Lemma~\ref{lem:lem4}) we will leverage a powerful matrix-valued Bernstein bound originally proven in~\cite{Rec09} (Theorem 3.2). We restate this theorem, slightly rephrased to better suit our notation.
\begin{thm}\label{thm::recht}\textsc{[Theorem 3.2~of~\cite{Rec09}]}
Let $M_1,M_2,\ldots,M_s$ be independent, zero-mean random matrices in $\R^{n \times n}$. Suppose $\max_{t \in [s]} \left\{\norm{\E\left(M_tM_t^T\right)},\norm{\E\left(M_t^TM_t\right)}\right\}\leq \rho^2$ and $\norm{M_t} \leq \gamma$ for all $t \in [s]$. Then, for any $\tau > 0$,
$$\norm{\frac{1}{s} \sum_{t=1}^s M_t} \leq \tau$$
holds, subject to a failure probability of at most
$$2n \exp\left(-\frac{s\tau^2/2}{\rho^2 + \gamma \tau/3}\right).$$
\end{thm}
In order to apply the above theorem, using the notation of Algorithm~$1$, we set $M_t = \frac{\widehat A_{i_t j_t}}{p_{i_t j_t}} e_{i_t}e_{j_t}^T - \widehat A$ for all $t \in [s]$ to obtain
\begin{equation}\label{eqn:mainexp}
\frac{1}{s} \sum_{t=1}^s M_t = \frac{1}{s} \sum_{t=1}^s \left[\frac{\widehat A_{i_t j_t}}{p_{i_t j_t}} e_{i_t}e_{j_t}^T - \widehat A\right] = \widetilde A - \widehat A.
\end{equation}
Let $\zeromtx_{n}$ denote the all-zeros matrix of size $n$. It is easy to argue that $\E \left(M_t\right) = \zeromtx_{n}$ for all $t \in [s]$. Indeed, if we consider that $\sum_{i,j=1}^n p_{ij}=1$ and $\widehat{A} = \sum_{i,j=1}^n \widehat A_{i j} e_{i}e_{j}^T$ we obtain
\[\E\left(M_t\right) = \sum_{i,j=1}^n p_{ij} \left(\frac{\widehat A_{i j}}{p_{i j}} e_{i}e_{j}^T - \widehat A\right) = \sum_{i,j=1}^n \widehat A_{i j} e_{i}e_{j}^T - \sum_{i,j=1}^n p_{ij} \widehat{A} = \zeromtx_{n}.\]
Our next lemma bounds $\norm{M_t}$ for all $t \in [s]$.
\begin{lem}\label{lem:lem2}
Using our notation, $\norm{M_t} \leq 4n\epsilon^{-1}\frobnorm{\widehat A}^2$ for all $t \in [s]$.
\end{lem}

\begin{Proof}
First, using the definition of $M_t$ and the fact that $p_{i_t j_t} = \widehat A_{i_tj_t}^2/\frobnorm{\widehat A}^2$,
$$\norm{M_t} = \norm{\frac{\widehat A_{i_t j_t}}{p_{i_t j_t}} e_{i_t}e_{j_t}^T - \widehat A} \leq \frac{\frobnorm{\widehat A}^2}{\abs{\widehat A_{i_tj_t}}}+\norm{\widehat A} \leq
\frac{2n\frobnorm{\widehat A}^2}{\epsilon}+\frobnorm{\widehat A}. $$
The last inequality follows since all entries of $\widehat A$ are at least $\epsilon/(2n)$ and the fact that $\norm{\widehat A} \leq \frobnorm{\widehat A}$. We can now assume that
\begin{equation}\label{eqn:assumption}
\frobnorm{\widehat A} \leq \frac{2n\frobnorm{\widehat A}^2}{\epsilon}
\end{equation}
to conclude the proof of the lemma. To justify our assumption in Eqn.~(\ref{eqn:assumption}), we note that if it is violated, then it must be the case that $\frobnorm{\widehat A} < \epsilon /(2n)$. If that were true, then all entries of $\widehat A$ would be equal to zero. (Recall that all entries of $\widehat A$ are either zero or, in absolute value, larger than $\epsilon/(2n)$.) Also, if $\widehat A$ were identically zero, then \textit{(i)} $\widetilde A$ would also be identically zero and, \textit{(ii)} all entries of $A$ would be at most $\epsilon/(2n)$. Thus, $$\norm{A-\widetilde A} = \norm{A} \leq \frobnorm{A} \leq \sqrt{n^2 \frac{\epsilon^2}{4n^2}}=\frac{\epsilon}{2}.$$
Thus, if the assumption of Eqn.~(\ref{eqn:assumption}) is not satisfied, the resulting all-zeros $\widetilde{A}$ still satisfies Theorem~\ref{thm::main}.
\end{Proof}

\noindent Our next step towards applying Theorem~\ref{thm::recht} involves bounding the spectral norm of the expectation of $M_tM_t^T$. The spectral norm of the expectation of $M_t^TM_t$ admits a similar analysis and the same bound and is omitted.
\begin{lem}\label{lem:lem3}
Using our notation, $\norm{\E\left(M_tM_t^T\right)} \leq n\frobnorm{\widehat A}^2$ for any $t \in [s]$.
\end{lem}

\begin{Proof}
We start by evaluating $\E\left(M_t M_t^T\right)$; recall that $p_{ij} = \widehat A_{ij}^2/\frobnorm{\widehat A}^2$:
\begin{eqnarray*}
\E\left(M_tM_t^T\right) &=& \E\left(\left(\frac{\widehat A_{i_tj_t}}{p_{i_tj_t}}e_{i_t}e_{j_t}^T-\widehat{A}\right)\left(\frac{\widehat A_{i_tj_t}}{p_{i_tj_t}}e_{j_t}e_{i_t}^T-\widehat A^T\right)\right)\\
&=& \sum_{i,j=1}^n p_{ij}\left(\frac{\widehat A_{ij}}{p_{ij}}e_{i}e_{j}^T-\widehat{A}\right)\left(\frac{\widehat A_{ij}}{p_{ij}}e_{j}e_{i}^T-\widehat A^T\right)\\
&=& \sum_{i,j=1}^n \left(\frac{\widehat A_{ij}^2}{p_{ij}}e_{i}e_{i}^T-\widehat A_{ij} \widehat A e_j e_i^T - \widehat A_{ij} e_i e_j^T \widehat A^T +p_{ij}\widehat A\widehat A^T\right)\\
&=& \frobnorm{\widehat A}^2 \sum_{i=1}^n m_i\cdot e_i e_i^T  -\sum_{j=1}^n \widehat Ae_j \sum_{i=1}^n \widehat A_{ij} e_i^T - \sum_{j=1}^n\left(\sum_{i=1}^n\widehat A_{ij} e_i\right) \left(\widehat A e_j\right)^T + \sum_{i,j=1}^n p_{ij}
\widehat A \widehat A^T,
\end{eqnarray*}
where $m_i$ is the number of non-zeroes of the $i$-th row of $\widehat{A}$.
We now simplify the above result using a few simple observations: $\sum_{i,j=1}^n p_{ij}=1$, $\widehat A e_j = \widehat A^{(j)}$, $\sum_{i=1}^n \widehat A_{ij}e_i = \widehat A^{(j)}$, and $\sum_{j=1}^n \widehat A^{(j)} \left(\widehat A^{(j)}\right)^T = \widehat A \widehat A^T$. Thus, we get
\begin{eqnarray*}
\E\left(M_tM_t^T\right) &=& \frobnorm{\widehat A}^2 \sum_{i=1}^n m_i\cdot e_i e_i^T  -\sum_{j=1}^n \widehat A^{(j)} \left(\widehat A^{(j)}\right)^T - \sum_{j=1}^n \widehat A^{(j)} \left(\widehat A^{(j)}\right)^T + \widehat A \widehat A^T\\
&=& \frobnorm{\widehat A}^2 \sum_{i=1}^n m_i\cdot e_i e_i^T  - \widehat A \widehat A^T.
\end{eqnarray*}
Since $0\leq m_i \leq n$ and using Weyl's inequality (Theorem~$4.3.1$ of \cite{book:matrix_analysis:HornJohnson}), which states that by adding a positive semi-definite matrix to a symmetric matrix all its eigenvalues will increase, we get that
\[ -\widehat{A}\widehat{A}^T \preceq \E\left(M_tM_t^T\right) \preceq n \frobnorm{\widehat{A}}^2 \Id_n.\]
Consequently $\norm{\E{\left(M_t M_t^T\right)}} = \max\left\{ \norm{\widehat{A}}^2, n\frobnorm{\widehat{A}}^2\right\} = n \frobnorm{\widehat{A}}^2$.
\end{Proof}

\noindent We can now apply Theorem~\ref{thm::recht} on Eqn.~(\ref{eqn:mainexp}) with $\tau = \epsilon/2$, $\gamma = 4n\epsilon^{-1}\frobnorm{\widehat A}^2$ (Lemma~\ref{lem:lem2}), and $\rho^2 = n\frobnorm{\widehat A}^2$ (Lemma~\ref{lem:lem3}) . Thus, we get that $\norm{\widehat A - \widetilde A} \leq \epsilon/2$ holds, subject to a failure probability of at most $$2n \exp\left(-\frac{\epsilon^2 s/8}{\left(1+4/6\right)n\frobnorm{\widehat A}^2}\right).$$
Bounding the failure probability by  $\delta$ and solving for $s$, we get that
\begin{equation*}
s \geq \frac{14}{\epsilon^2}n\frobnorm{\widehat A}^2 \ln\left(\frac{2n}{\delta}\right).
\end{equation*}
Using $\frobnorm{\widehat A} \leq \frobnorm{A}$ (by construction) concludes the proof of the following lemma, which is the main result of this section.
\begin{lem}\label{lem:lem4}
Using the notation of Algorithm~$1$, if
$s \geq 14n\epsilon^{-2} \frobnorm{A}^2\ln\left(2n/\delta\right),$
then, with probability at least $1-\delta$, $$\norm{\widehat A - \widetilde A} \leq \epsilon/2.$$
\end{lem}

\section{Acknowledgments}

We would like to thank the anonymous reviewers for numerous comments that significantly improved the presentation of our work. This research has been supported by the National Science Foundation through NSF CCF 1016501, NSF DMS 1008983, and NSF CCF 545538 awards to Petros Drineas.
{\small
\bibliographystyle{alpha}

}
\end{document}